# Charge and hydration structure of dendritic polyelectrolytes: molecular simulations of polyglycerol sulphate†

Rohit Nikam,[ab] Xiao Xu,[c] Matthias Ballauff,[bde] Matej Kanduč [*a] and Joachim Dzubiella[*af]

Macromolecules based on dendritic or hyperbranched polyelectrolytes have been emerging as high potential candidates for biomedical applications. Here we study the charge and solvation structure of dendritic polyglycerol sulphate (dPGS) of generations 0 to 3 in aqueous sodium chloride solution by explicit-solvent molecular dynamics computer simulations. We characterize dPGS by calculating several important properties such as relevant dPGS radii, molecular distributions, the solvent accessible surface area, and the partial molecular volume. In particular, as the dPGS exhibits high charge renormalization effects, we address the challenges of how to obtain a well-defined effective charge and surface potential of dPGS for practical applications. We compare implicit- and explicit-solvent approaches in our all-atom simulations with the coarse-grained simulations from our previous work. We find consistent values for the effective electrostatic size (*i.e.*, the location of the effective charge of a Debye–Hückel sphere) within all the approaches, deviating at most by the size of a water molecule. Finally, the excess chemical potential of water insertion into dPGS and its thermodynamic signature are presented and rationalized.

## 1 Introduction

Macromolecules based on dendritic or hyperbranched polyelectrolytes have attracted strong interest from scientists in recent years due to their versatile bioapplications, such as drug delivery, tissue engineering, and biological imaging.[1–3] High potential candidates for use in medical treatments have been identified as hyperbranched or dendritic polyelectrolytes, such as polyglycerol sulphate (hPGS or dPGS).[4] The latter have been found to be very efficient for the treatment of neurological disorders arising from inflammation,[5] as therapeutics for the prevention of tissue damage,[6] as substance delivery platforms[7,8] (*e.g.*, for transporting drugs to tumour cells[9]) and as imaging agents for the diagnosis of rheumatoid arthritis.[8] Due to their charged terminal groups, dPGS particles interact mainly through electrostatic effects.[10] The high anionic surface charge is therefore the basis for dPGS's high anti-inflammatory potential through binding to relevant proteins.[11,12]

The important applications of dendritic macromolecules in general have initiated substantial efforts towards their detailed molecular-level characterization by theory and computer simulations.[1,3] A large number of fully atomistic computer simulations, for example, of poly(amidoamine) (PAMAM) based dendrimers have been performed,[13–19] some of them with particular focus on explicit solvent effects.[20–27] On the other hand, to overcome the limitation of the system size of atomistic simulations, coarse-grained (CG) monomer-resolved models, which contain various levels of specific chemical features, have led to plentiful structural insight on larger scales.[28–53] For the case of (internally or surface) charged dendrimers, one important focus has been on the dominant role of condensed counterions and charge renormalization[54–61] in modulating the conformation and effective charge of the dendrimers.[49–53] Indeed, for highly charged polyelectrolytes, electrostatic effects naturally dominate the interactions with proteins and have complex dependencies on the effective size, charge, flexibility, and the shape and charge heterogeneity of the interaction partners.[10,62–67]

Despite the large body of studies on dendrimers, apart from a notable exception,[19] it has been hardly attempted to define and determine the effective surface potential (and its location) of charged dendrimers so far, despite its significance for the electrostatic interaction with macromolecules or the response to electric fields. The challenges are the integration of the very heterogeneous and long-ranged charge distributions of all constituents as well as finding a reasonable or at least practical definition for the surface potential and its precise location. In our very recent work we thus thoroughly reconsidered and investigated the key electrostatic features of the charged dendrimers for the case of dPGS using coarse-grained computer simulations.[68] In these implicit-solvent/explicit-salt Langevin dynamics simulations, we studied dPGS up to its sixth generation. We argued that a systematic mapping of the long-range decay of the calculated electrostatic potentials onto the Debye–Hückel form for simple charged spheres serves as the most practical defining equation for the effective electrostatic properties seen in the far field. Hence, this led to the determination of well-defined effective net charges and corresponding radii, surface potentials, and surface charge densities of dPGS. The latter were found to be up to one order of magnitude smaller than the bare values, consistent with previously derived theories on charge renormalization and weak saturation for high dendrimer generations (charges). The surface potentials of dPGS were found to agree with electrophoretic experiments, while still some tolerance in the comparison had to be imposed to leave room for hydration effects.[68]

*a Research Group Simulations of Energy Materials, Helmholtz-Zentrum Berlin für Materialien und Energie, Hahn-Meitner-Platz 1, D-14109 Berlin, Germany. E-mail: joachim.dzubiella@helmholtz-berlin.de, matej.kanduc@helmholtz-berlin.de*
*b Institut für Physik, Humboldt-Universität zu Berlin, Newtonstr. 15, D-12489 Berlin, Germany*
*c School of Chemical Engineering, Nanjing University of Science and Technology, 200 Xiao Ling Wei, Nanjing 210094, P. R. China*
*d Soft Matter and Functional Materials, Helmholtz-Zentrum Berlin für Materialien und Energie, Hahn-Meitner-Platz 1, D-14109 Berlin, Germany*
*e Multifunctional Biomaterials for Medicine, Helmholtz Virtual Institute, Kantstr. 55, D-14513 Teltow-Seehof, Germany*
*f Physikalisches Institut, Albert-Ludwigs-Universität Freiburg, Hermann-Herder Str. 3, D-79104 Freiburg, Germany. E-mail: joachim.dzubiella@physik.uni-freiburg.de*
† Electronic supplementary information (ESI) available. See DOI: 10.1039/c8sm00714d

The latter work was based on a coarse-grained force field where the explicit action of water was neglected and the charged atoms were clumped together in beads. Then always the questions remains, how do these results compare to fully resolved, explicit-water simulations and what are the details of the water structural effects?[27] Here we address this question with a focus on electrostatic properties, and revisit the electrostatic dendrimer problem with a fully atomistic representation of dPGS of generations 0 to 3 in explicit-water and electrolyte (NaCl) solution, cf. Fig. 1. The inclusion of water gives rise to larger complexity of the problem, in particular due to explicit and local screening effects, which are absent in the implicit-solvent simulations. We re-address the challenges of how to obtain a well-defined effective charge and surface potential of dPGS for practical applications and compare coarse-grained, implicit-, and explicit-solvent approaches. We eventually find consistent values for the effective electrostatic size (i.e., the location of the effective charge of a Debye–Hückel sphere) within all the approaches, including the most coarse-grained,[68] deviating at most by the size of a water molecule. In addition, we seize the opportunity and take a closer look at the thermodynamics of water insertion.[21,69,70] Water insertion and release into and from the penetrable dendrimer may lead to significant contributions to the thermodynamic signatures of binding of the dendrimers to proteins.[71]

## 2 Simulation details

Initial atomistic structures of all generations ($n$ = 0–3) of the PGS dendrimers [cf. Fig. 1(b)] were constructed using the Marvin software [Marvin 16.4.11.0, 2016, ChemAxon (www.chemaxon.com)]. $Na^+$ and $Cl^-$ ions were used as counterions and coions, respectively. MD simulations were performed using the GROMACS 5.0.6 platform,[72–75] employing the General Amber Force Field (GAFF)[76,77] for dPGS and ions. Partial charges for dPGS atoms were calculated using the AM1-BCC quantum mechanical charge model,[78] which is compatible with GAFF. The Antechamber package[77,79] from USCF Chimera software[80] was used to assign the partial charges, which are summarized along with the force-field parameters in the ESI.†

The structures were subject to a series of initialization and equilibration protocols. First, the dendrimer was immersed in a cubic simulation box of SPC/E water.[81] Pertaining to the aim of studying the electrostatic potential distribution around the dPGS dendrimer and in order to curb finite size effects, large sizes of the boxes were used ranging from 13 nm for $G_0$-dPGS to 26 nm for $G_3$-dPGS (cf. Table 1). We then inserted appropriate numbers of $Na^+$ and $Cl^-$ ions to ensure electroneutrality and a

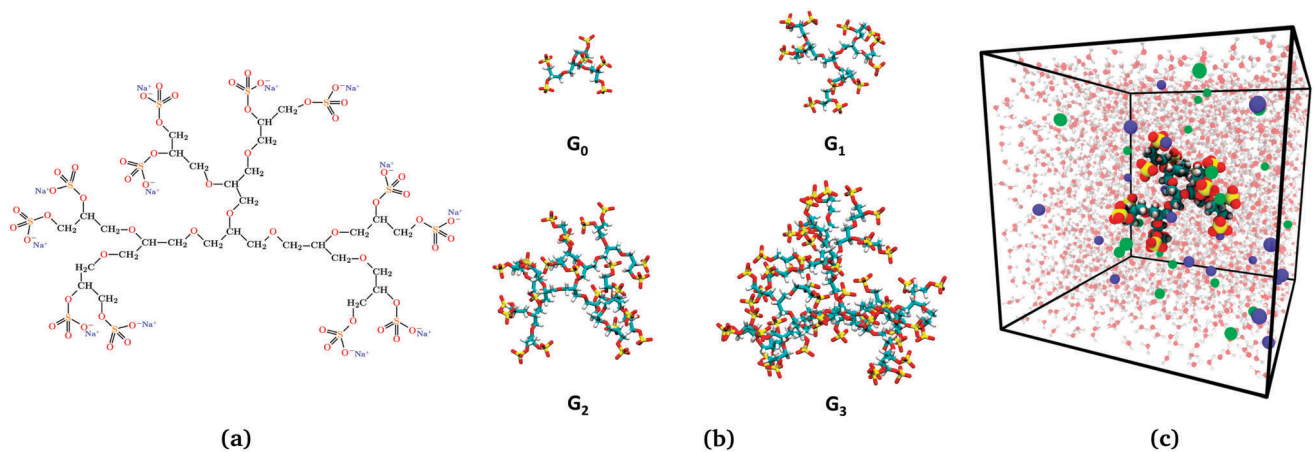

Fig. 1 (a) Illustration of the dendritic PGS (dPGS) with an example of the chemical structure of $G_1$-dPGS along with the counterions ($Na^+$) of the respective sulphate groups. (b) Simulation snapshots of all four dPGS generations. (c) Simulation snapshot of the simulation box of 15.7 nm size, showing $G_1$-dPGS with $Na^+$ (green spheres) and $Cl^-$ (blue spheres) ions acting as counter- and coions, respectively. Water is also shown in the background (not to scale).

Table 1  System details for the simulations of the four different generations. The pressure and temperature are fixed to 1 bar and 300 K, respectively, and the bulk salt concentration is 25 mM

| Generation | Number of dPGS atoms | Number of water molecules | Box length (nm) | Number of counterions (Na$^+$) | Number of coions (Cl$^-$) |
| --- | --- | --- | --- | --- | --- |
| 0 | 65 | 66 394 | 12.6 | 32 | 26 |
| 1 | 149 | 128 075 | 15.7 | 71 | 59 |
| 2 | 317 | 308 645 | 21.0 | 163 | 139 |
| 3 | 653 | 583 725 | 25.9 | 313 | 265 |

bulk salt concentration of 25 mM. The system was then subject to energy minimization and 100 ns equilibration in the NPT ensemble at 1 bar and 300 K. The production MD simulations in the same NPT ensemble for each of the dendrimers were subsequently performed for 150 ns. Bonds involving hydrogens were constrained by the LINCS algorithm.[82] The electrostatic interactions were calculated with the Particle Mesh Ewald method[83] using 1 nm as a cut-off for the short-range part. The Lennard-Jones cut-off was also set to 1 nm. The details of the simulation conditions for all generations are summarised in Table 1.

## 3 Analysis methods

### 3.1 Potential of mean force and the structure of the electrolyte

We describe and analyse the structure of the electrolyte solution surrounding the dPGS by calculating the radial distribution functions (RDFs) between the dPGS-COM (center-of-mass) and ions/water, $g_i(r)$ (i = Na$^+$/Cl$^-$/water). Performing a Boltzmann inversion[84,85] of $g_i(r)$ gives the corresponding potential of mean force (PMF) as follows

$$\beta V_i(r) = -\ln[g_i(r)] \quad (1)$$

Owing to the charged system in the present case, the PMF of the ions (or RDF) $V_\pm(r)$ (where $\pm$ stands for Na$^+$/Cl$^-$) can be decomposed into short-range and long-range contributions as[86-88]

$$V_\pm(r) = V_\pm^{sr}(r) + V_\pm^{lr}(r). \quad (2)$$

The long-range part $V_\pm^{lr}(r)$ can be typically approximated by a Debye–Hückel (DH) type of potential $\phi_{DH}(r)$[87,89] of the form

$$V_\pm^{DH}(r) = \pm e_0 \phi_{DH}(r) \quad (3)$$

The basic DH model[89] for the radial electrostatic potential around a charged sphere with the radius $r_{eff}$ and charge (valency) $Z_{eff}$ is

$$e_0 \beta \phi_{DH}(r) = Z_{eff} l_B \frac{e^{\kappa r_{eff}}}{1 + \kappa r_{eff}} \frac{e^{-\kappa r}}{r} \quad (4)$$

where $e_0$ is the elementary charge, $\beta^{-1} = k_B T$ is the thermal energy, $\kappa = \sqrt{8\pi l_B \rho_0}$ is the inverse Debye length for a symmetric monovalent salt of bulk concentration $\rho_0$ and $l_B = \beta e_0^2 / 4\pi \varepsilon_0 \varepsilon_r$ is the electrostatic coupling parameter called the Bjerrum length, which is 0.72 nm in water at room temperature. This solution is derived using two boundary conditions, i.e., fixing the effective surface charge $Z_{eff}$ and the potential far away, $\phi(\infty) = 0$. The short-range part $V_\pm^{sr}(r)$ in eqn (2) includes all non-DH effects due to specific dPGS–ion interactions and ion–ion correlations.

Eqn (3) can now be rewritten as[90]

$$\ln|\beta V_\pm^{DH}(r)r| = \ln\left|\pm Z_{eff} l_B \frac{e^{\kappa r_{eff}}}{1 + \kappa r_{eff}}\right| - \kappa r \quad (5)$$

where the right-hand side turns out to be a linear function with a negative slope defined by the inverse Debye length $\kappa$. This construction was used before[90] to calculate the short-range part of specific ion–ion interactions by subtracting the linear DH fit from the full PMF. In our work it will also serve as a method to identify the location where the linear long-range DH decay (i.e., the diffusive double layer) crosses over to the non-linear electrostatic behaviour of the correlated and condensed ions deep in the surface layer. This should be in principle one possible reasonable definition for an effective size and charge in the DH picture, $r_{eff}$ and $Z_{eff}$ in eqn (4), of the dPGS with respect to its charge properties.

### 3.2 Charge distribution and surface potentials

Instead of mapping to ionic PMFs, we showed before that equivalently a direct mapping of the long-range decay of the total electrostatic potential onto the spherical DH form may also serve as a practical defining equation for the effective properties seen in the far field.[68] In general, the local, radially symmetric electrostatic potential $\phi(r)$ can be in principle directly evaluated from the radial charge density distributions as a function of the distance from the center of mass (COM) of dPGS through Poisson's equation

$$\nabla^2 \phi(r) = -\sum_i \frac{Z_i \rho_i(r)}{\varepsilon_0 \varepsilon_r} \quad (6)$$

Here, the summation runs over all atomic species $i$ (dPGS atoms, counter- and coions, as well as the hydrogen and oxygen atoms of water) with a non-zero partial charge. Here, $\rho_i(r)$ is the radial number density distribution and $Z_i$ is the corresponding valency of the partial charge. The potential $\phi(r)$ is obtained by integrating eqn (6) twice. In addition to the potential profile, we calculate the running coordination number defined as,

$$N_i(r) = \int_0^r \rho_i(r') 4\pi r'^2 dr' \quad (7)$$

This motivates the definition of total cumulative charge as

$$Z(r) = \sum_i Z_i N_i(r) \quad (8)$$

We aim to define the effective charge of the dendrimer $Z_{eff}$ and the effective surface potential $\phi_0$ within the linear DH picture.

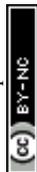

The DH theory works well in the far-field limit whereas the short-range non-linear effects arising from condensed counterions are neglected and absorbed in the effective charge (as demonstrated, e.g., by solutions of the full non-linear Poisson–Boltzmann theory[91–95]). Analogous to the DH constructions of the PMFs above, by mapping the logarithm of eqn (4) directly on the far-field behaviour of the electrostatic potential profile obtained from simulations, the diffusive double-layer behaviour and $r_{eff}$ and $Z_{eff}$ can be quantified with high accuracy.[68] The effective DH surface potential, as seen in the far field, is then simply defined as $\phi_0 = \phi_{DH}(r_{eff})$.

### 3.3 Inflection point criterion

It is also customary, for highly charged macromolecules, to define an effective radius as an inflection point in the plot of $Z$ vs. the inverse radial distance $1/r$.[68,94,96] It follows from the Poisson–Boltzmann and counterion-condensation theory that the condition $d^2Z/d(1/r)^2|_{r=r_{inf}} = 0$ leads to a radius $r_{inf}$, within which ions are assumed to be *condensed* and as such serves also as a defining radius to separate the linear (double-layer) DH regime from the non-linear regime.

### 3.4 Implicit vs. explicit-water integration

In order to better scrutinize the explicit water effects, we attempt to compare two approaches for calculating the radial electrostatic potential and the total cumulative charge using the charge integration and mapping as described in Section 3.2. The comparison is based on the inclusion/exclusion of the partial charges of water in eqn (6) as follows:

• "Explicit water" approach: Here we include the partial charges of all the atoms in the simulation box, which include dPGS atoms, counterions, coions and the partial charges of the water molecules. This basically assumes a multi-ingredient mixture of different charged species in vacuum, thus taking the dielectric constant of unity while calculating the electrostatic potential.

• "Implicit water" approach: This approach assumes an implicit water model for the integration, that is, for calculating the electrostatic potential in eqn (6) the partial charges of water are excluded; thereby the dielectric constant of the medium is chosen as 72 (corresponding to the SPC/E water model and $T = 300$ K, ref. 97). The charge density distributions of all other species (i.e., dPGS atoms, counterions, and coions) are deployed to calculate the potential. This approach should be valid for the potential in the far-field regime where the dielectric constant, to a good approximation, is determined by the bulk solvent.

Note that the DH construction described in Section 3.1 is essentially an explicit-water approach as it directly works on the ionic profiles, not the integrated potentials. Therefore, no approximation is made concerning the dielectric properties as in the implicit-water approach to the potential.

### 3.5 Gibbs dividing surface and partial molecular volume

The description of the dPGS–water interface in terms of the radial density distribution of water around the dPGS-COM allows us to consider a phase dividing surface between adjoining water phases in the dPGS and bulk liquid. Such an interface is typically defined by the Gibbs dividing surface[98–103] (GDS). Assuming its spherical nature, it is given by[104]

$$\frac{4\pi}{3} r_{GDS}^3 = K_w, \quad (9)$$

where the Kirkwood–Buff integral[105] $K_w$ defines the partial molecular volume of dPGS, $\bar{V}_d$, given by

$$K_w = \bar{V}_d = \int_0^\infty [1 - g_w(r)] 4\pi r^2 dr \quad (10)$$

where $g_w(r)$ is the radial distribution function of water molecules with respect to the dPGS-COM. Eqn (10) essentially describes the effective volume occupied by one dPGS molecule. The number of replaced water molecules is then given by $\Delta n_w = \rho_w^0 \bar{V}_d$, with $\rho_w^0$ being the water bulk density.

### 3.6 Water penetration thermodynamics

The thermodynamic signature of the dPGS–water association can be obtained from the temperature-dependence of the PMF, thereby obtaining distance-resolved profiles of free energy, enthalpy and entropy at the atomistic level.[70,106–108] For this we identify the distance-resolved PMF, $V_w(r)$, as the dPGS–water (Gibbs) free energy of association, $G_w(r)$. From the latter the corresponding entropy can be determined *via* the standard thermodynamic equation,

$$S_w(r) = -\left(\frac{\partial G_w(T,r)}{\partial T}\right)_{N,P} \quad (11)$$

The corresponding distance-resolved water enthalpy is then

$$H_w(r) = G_w(r) + TS_w(r) \quad (12)$$

## 4 Results and discussion

### 4.1 Density distribution functions

**4.1.1 Terminal groups.** Fig. 2 shows the atomic radial density distributions with respect to the dPGS-COM. The radial density of the terminal sulphur is inhomogeneous and has pronounced peak(s) as shown in Fig. 2(a). The growing branching leads to shifts of the positions of the maxima to larger distances with increasing generation. A single peaked distribution is found for generations 0 and 1 indicating that most of the terminal sulphur atoms stay on the molecular surface. The breadth of distribution increases with generation. For generations 2 and 3, the distribution becomes bimodal with a minor peak at $r = 0.6$ nm. This indicates that some of the sulphate groups reside in the interior of the dendrimer. This penetration of terminal groups into the interior volume, called "backfolding", has already been observed in previous studies.[109–112] The increase in the number of terminal sulphate groups with higher generation leads to higher charge density in the dPGS corona,[68] leading to higher electrostatic repulsion that essentially leads to backfolding. We use the major peak position of the distribution to define the "structural" or "intrinsic" dPGS radius $r_d$ (*cf.* Table 2). We observe a monotonic increase in $r_d$ with increasing generation from $r_d = 0.66$ nm for G$_0$ to $r_d$ = 1.4 nm for G$_3$. We also compare these quantities to the standard radius of gyration in Table 2.

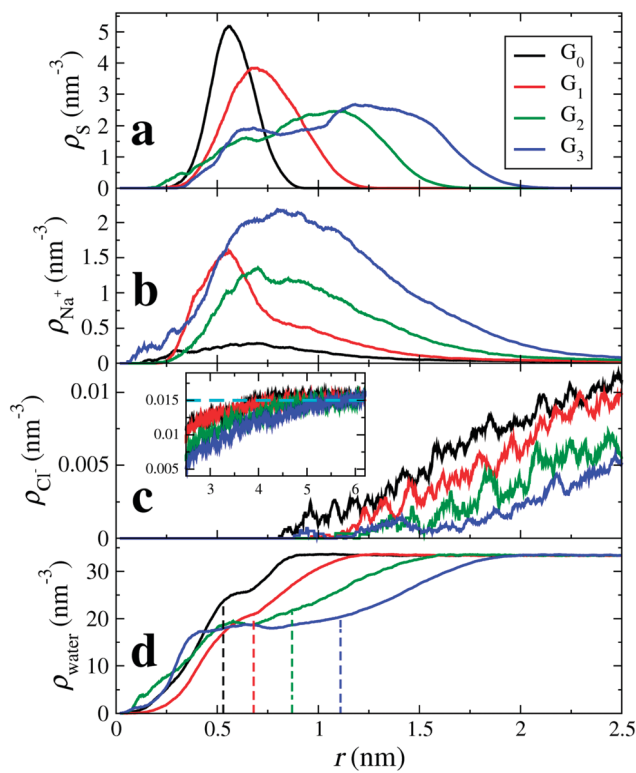

Fig. 2 Radial density distributions $\rho_i(r)$ with respect to the dPGS-COM of (a) the sulphur atoms of the terminal sulphate groups, i = S, (b) cations (counterions; i = Na$^+$), (c) anions (coions; i = Cl$^-$), and (d) water molecules (i = water) for all four generations. In the inset of panel (c) the anion density profiles at larger distances are shown, with the bulk density $\rho_0$ = 25 mM marked by a dashed horizontal line. The vertical dashed lines indicate the location of the dPGS–water Gibbs dividing surface $r_{GDS}$ for each generation (cf. Table 3).

With the knowledge of the intrinsic dPGS radius $r_d$, along with the bare charge $Z_0$, we can define the dPGS bare surface charge density $\sigma_d = Z/4\pi r_d^2$. The number of terminal sulphate groups increases with increasing generation, thus increasing $\sigma_d$. On the other hand, the tendency of backfolding of the terminal groups also increases with increasing generation, thus hampering the growth of $\sigma_d$. Table 2 shows that the net result is a monotonic increase in $\sigma_d$, implying a minor contribution of backfolding.

**4.1.2 Counterions and salt.** Fig. 2(b) shows the counterion density distributions, which exhibit a single peak and decay in the exponential (DH-like or Yukawa) fashion for $r \gtrsim r_d$, while reaching bulk concentration at large distances. Whereas the electrostatic attraction of the terminal sulphate groups drives counterions towards dPGS, excluded volume restricts their entropy. This combined effect leads to the non-monotonic distribution. The coions, on the other hand, as indicated in Fig. 2(c), are depleted from the dPGS interior due to electrostatic repulsion.

**4.1.3 Water.** Apart from the open morphology of dPGS, which is favourable for the water uptake, the electrostatic repulsion between like-charged terminal sulphate groups and their polar nature facilitate their interaction with water. Hence, water penetrates partially into the interior of dPGS as depicted in Fig. 2(d), which shows the radial density of water as a function of the distance from the dPGS-COM for all generations. The water density gradually rises as we go radially outward from the dPGS-COM and reaches its bulk value. We see that besides the water penetration into the dendrimer interior, the density profile starts to develop a peak with increasing generation at the region around 0.6 nm. This could be attributed to the backfolding of sulphate groups in the interior region of the dendrimer, since water indulges in a preferential interaction with sulphate groups due to their polar nature. The dashed lines represent the locations of the Gibbs dividing surfaces $r_{GDS}$ of the individual generations reported in Table 3. It can be seen that $r_{GDS}$ increases almost linearly with increasing generation and is roughly 80% of the bare radius $r_d$ for all generations (cf. Tables 2 and 3). As a net effect, water is excluded and the partial molecular volume is positive, cf. Table 3. From generation 0 to 3, ca. $\Delta n_w$ = 21 to 198 water molecules are replaced by dPGS, respectively.

Another effective means to characterize the dendrimer hydration properties is by calculating the solvent accessible surface area (SASA). Here, each dendrimer atom is assumed as a sphere with the radius $r_i$ being the sum of the van der Waals radius of that atom, $r_{vdW}$, and a water 'probe' radius $r_p$, i.e., $r_i = r_{vdW} + r_p$. The dendrimer is thus assumed as a union of such fused spheres. The SASA is defined as the surface traced by the spherical solvent probe as it rolls around the van der Waals spheres of the dendrimer.[20] The values of the SASA for a typical probe radius of 0.15 nm for all generations are listed in Table 3. More details and results using other probe radii can be found in the ESI.†

Table 2 Some basic structural parameters of the simulated dPGS. MW is the molecular weight of the dPGS while $r_d$ and $Z_0$ stand for the intrinsic radius (defined by the terminal sulphur peak position in the density distribution) [cf. Fig. 2(a)] and bare charge, respectively, leading to the bare dPGS surface charge density $\sigma_d$. $R_g$ is the radius of gyration

|  | G$_0$ | G$_1$ | G$_2$ | G$_3$ |
|---|---|---|---|---|
| MW (kDa) | 0.79 | 1.72 | 4.10 | 8.32 |
| $r_d$ (nm) | 0.66 | 0.86 | 1.16 | 1.40 |
| $Z_0$ ($e_0$) | −6 | −12 | −24 | −48 |
| $\sigma_d$ ($e_0$ nm$^{-2}$) | −1.08 | −1.30 | −1.41 | −1.93 |
| $R_g$ (nm) | 0.57 | 0.75 | 1.03 | 1.28 |

Table 3 Some hydration properties of the simulated dPGS. $r_{GDS}$ stands for the Gibbs dividing surface. SASA is the solvent accessible surface area of dPGS evaluated with the probe radius of 0.15 nm, which is approximately the radius of one water molecule.[113] $\bar{V}_d$ is the partial molecular volume of a single dPGS and $\Delta n_w$ denotes the corresponding number of water molecules replaced by dPGS

|  | G$_0$ | G$_1$ | G$_2$ | G$_3$ |
|---|---|---|---|---|
| $r_{GDS}$ (nm) | 0.53 | 0.69 | 0.88 | 1.12 |
| SASA (nm$^2$) | 9.56 | 16.60 | 34.38 | 61.16 |
| $\bar{V}_d$ (nm$^3$) | 0.64 | 1.34 | 2.83 | 5.88 |
| $\Delta n_w$ | 21.24 | 44.35 | 91.32 | 197.55 |

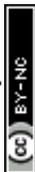

### 4.2 Electrostatic properties of dPGS

**4.2.1 Effective size and charge I: DH construction on counter-ion profiles.** First, we define the effective size and charge of dPGS by adopting the method described in Section 3.1, *i.e.*, we quantitatively map the long-range part of the dPGS–counterion PMF $V_+(r)$ obtained by the Boltzmann inversion of the counterion radial number distribution $g_{Na+}(r)$ onto the basic DH theory [eqn (4)]. Note again that here no assumptions have to be made on the dielectric constant of water. Fig. 3 shows the logarithm of a rescaled PMF, $r\beta V_+(r)$, as a function of distance from the dPGS-COM for all generations. We see that at large distances, the profiles decay linearly for all generations with a slope $\kappa = 0.52$ nm$^{-1}$ consistently corresponding to the salt concentration $\rho_0 = 25$ mM. As opposed to the exponential behaviour at large distances, the PMF reaches a maximum at smaller distances before it decreases to almost zero close to the dPGS core. This highly non-linear behaviour is expected and can be attributed to high electrostatic and steric correlations between the dPGS atoms and counterions. As shown in Fig. 3, we can now set the boundary between the long-range ($r > r_{eff}$) DH-like PMF $V^{DH}(r)$, which decays linearly and the short-range ($r_d < r < r_{eff}$) PMF, which is non-linear. As a criterion, we define $r_{eff}$ as the shortest distance where the DH fit (minimizing the root mean square deviation to the PMF) and PMF cross before the maximum of the PMF.[68] This formulation treats the dPGS as a homogeneously charged sphere of an effective radius $r_{eff}$ at which the effective surface charge $Z_{eff}$ can be defined. Fig. 3 shows the location of $r_{eff}$ for each generation with vertical dotted lines. The corresponding electrostatic properties are listed in Table 4.

**4.2.2 Effective size and charge of dPGS II: DH construction using the electrostatic potential (implicit water).** The radial density distributions of charges can be now utilised to calculate the local charge accumulation (or running ion coordination) and electrostatic potential distribution around dPGS. In the "implicit water" approach, the total cumulative charge $Z$, or in other words the local net charge [according to eqn (8)] is shown in Fig. 4(a). Far away from the dPGS-COM, the charge build-up due to the dPGS terminal groups leads to a more negative $Z$ close to dPGS. Subsequently, we see a reversal in its profile at a distance where the counterion accumulation starts to become dominant and the magnitude of $Z$ tends to decrease onwards. This so-called charge renormalization effect has been extensively studied and a wide variety of theories have been developed for the effective charge and size of simple charged spheres with smooth surfaces.[91,94–96,114] A comparison of the cumulative charge $Z$ distributions of the implicit *versus* explicit-water integrations agrees in their long-range decay. The explicit-route profiles are very noisy due to strong water fluctuations (see ESI†). A comparison of implicit and explicit approaches on the level of the electrostatic potential will be discussed later.

Xu *et al.*[68] in their coarse-grained simulations defined the effective size and charge by the electrostatic potential mapping procedure described above in Section 3.2. Adopting this implicit approach and assuming a dielectric constant of $\varepsilon_r = 72$, we plot in Fig. 4(b) the logarithm of the rescaled potential $|re_0\beta\phi(r)|$ as a function of distance from the dPGS-COM for all generations. At large distances, the profiles decay linearly for all generations with the slope $\kappa = 0.52$ nm$^{-1}$, corresponding consistently to the salt concentration of $\rho_0 = 25$ mM. Hence, it is possible in this approach to define the effective electrostatic surface potential $\phi_0$ as the potential at $r_{eff}$ [$\phi_0 = \phi(r_{eff}) = \phi_{DH}(r_{eff})$]. As before, we define $r_{eff}$ as the shortest distance where the DH fit (minimizing the root mean square deviation to the PMF) and PMF cross before the maximum of the PMF.[68] The location of $r_{eff}$ for each generation is shown in Fig. 4 by vertical dotted lines. The intersection points in Fig. 4(a and c) denote $Z_{eff}$ and the number of cumulative counterions at $r_{eff}$, $N_C(r_{eff})$, respectively. We see that $r_{eff}$ increases with increasing generation. On comparing $Z_{eff}$ values with the corresponding bare charges $Z_0$ (*cf.* Tables 2 and 4), a significant charge renormalization can be observed.

**4.2.3 Effective size and charge III: inflection point criterion.** Within counterion-condensation theory one can also define the crossover radius from the diffusive to the condensed ionic regimes as an inflection point in the plot of $Z$ *vs.* the inverse radial distance $1/r$.[68,94,96] Inflection points are shown in Fig. 4 as empty circle symbols. The corresponding effective potential and the number of condensed counterions can be read from circle symbols at the respective vertical axes in Fig. 4(b and c), respectively.

**4.2.4 Discussion.** Table 4 summarizes the electrostatic attributes of dPGS stemming from their different definitions as we just discussed as well as from the coarse-grained simulations of Xu *et al.*[68] Overall, all approaches to $r_{eff}$ and $Z_{eff}$ agree reasonably well; the values of $r_{eff}$ deviate from each other by at most the size of a water molecule $\simeq 0.3$ nm. More surprisingly, despite the large charge renormalization, the values of $Z_{eff}$ from different approaches deviate only by at most three elementary charges (for the largest generation where the bare charge is $-48$ $e_0$).

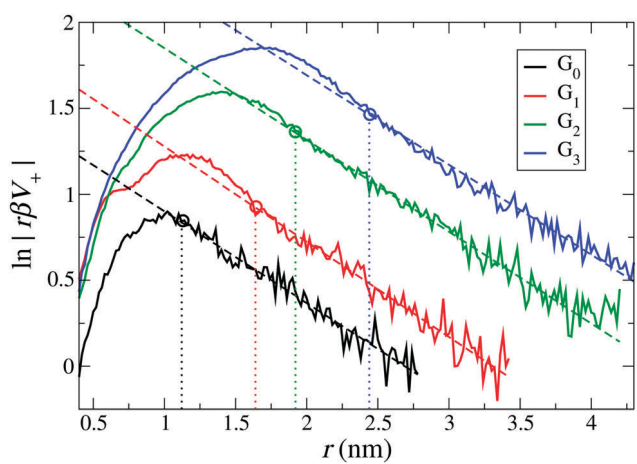

Fig. 3 The rescaled PMF $V_+(r)$ between dPGS and Na$^+$ ions plotted for all generations. The dashed lines are fits of eqn (5) in the far-field. Their slopes determine the inverse Debye length $\kappa = 0.52$ nm$^{-1}$, which corresponds to the salt concentration $\rho_0 = 25$ mM. The dotted vertical lines indicate the effective radius of dPGS, $r_{eff}$, which separates the non-linear short-range nature of the electrostatic regime from the long-range linear DH regime (see also the text). The corresponding values of the effective dPGS charge $Z_{eff}$, the number of bound counterions $N_C$, and the effective charge density $\sigma_{eff}$ are summarized in Table 4.

Table 4 Electrostatic parameters of dPGS obtained using constructions from Kalcher et al.[90] (I. counterion density route), Xu et al.[68] (II. Implicit-water potential route) and the inflection point criterion[68,94,96] (III) evaluated at a salt concentration of $\rho_0$ = 25 mM. Values from previous coarse-grained simulations by Xu et al.[68] are also compared. Here, $r_{eff}$ is the effective dPGS radius. $Z_{eff}$ stands for the effective dPGS charge due to the charge renormalization by the condensed counterions. $\sigma_{eff}$ thus is the effective surface charge density while $\phi(r_{eff})$ is the effective electrostatic potential at $r_{eff}$

| Method | Label | $G_0$ | $G_1$ | $G_2$ | $G_3$ |
|---|---|---|---|---|---|
| I. Counterion density route | $r_{eff}$ (nm) | 1.05 | 1.63 | 1.94 | 2.47 |
| | $Z_{eff}$ ($e_0$) | −4.98 | −6.20 | −9.59 | −11.39 |
| | $\sigma_{eff}$ ($e_0$ nm$^{-2}$) | −0.36 | −0.18 | −0.20 | −0.14 |
| II. Potential route (implicit) | $r_{eff}$ (nm) | 1.15 | 1.70 | 2.10 | 2.57 |
| | $Z_{eff}$ ($e_0$) | −4.80 | −6.06 | −8.95 | −10.96 |
| | $\sigma_{eff}$ ($e_0$ nm$^{-2}$) | −0.29 | −0.17 | −0.16 | −0.13 |
| | $\phi(r_{eff})$ ($k_B T$) | −1.56 | −1.30 | −1.78 | −1.52 |
| III. Inflection point | $r_{inf}$ (nm) | 1.09 | 1.43 | 1.94 | 2.36 |
| | $Z_{inf}$ ($e_0$) | −4.90 | −6.80 | −9.67 | −12.27 |
| | $\sigma_{inf}$ ($e_0$ nm$^{-2}$) | −0.33 | −0.26 | −0.20 | −0.17 |
| | $\phi(r_{inf})$ ($k_B T$) | −1.72 | −1.82 | −2.12 | −1.86 |
| CG simulation (Xu et al., Ref. 68) potential route ($\rho_0$ = 10 mM) | $r_{eff}$ (nm) | 0.70 | 1.60 | 1.90 | 2.40 |
| | $Z_{eff}$ ($e_0$) | −6.00 | −7.30 | −10.60 | −14.30 |
| | $\sigma_{eff}$ ($e_0$ nm$^{-2}$) | −0.97 | −0.23 | −0.23 | −0.20 |
| | $\phi(r_{eff})$ ($k_B T$) | −4.20 | −2.12 | −2.37 | −2.22 |

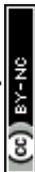

In all approaches we see an increase in $r_{eff}$ and the magnitude of $Z_{eff}$ with increasing dPGS generation along with a substantial charge renormalization. As an example, $G_3$-dPGS has a bare charge $Z$ of −48 $e_0$ (cf. Table 2), which is effectively renormalized to $Z_{eff} \sim -11$ $e_0$ seen at larger distances $r > r_{eff}$ according to the DH mapping of the potential (Method II). $|Z_{eff}|$ increases with the dPGS generation, but in a much weaker fashion than the bare charge, demonstrating that the counterion condensation effect strengthens with growing generation. We observe a consistent shift of the inflection point away from the dPGS as shown in Fig. 4(b and c).

Tables 2 and 4 show that the effective surface charge density $\sigma_{eff} = Z_{eff}/4\pi r_{eff}^2$ is about an order of magnitude lower than the bare one $\sigma_d$. Interestingly, the values of $\sigma_{eff}$ decrease with increasing generation as opposed to the $\sigma_d$ values. Experiments on carboxyl-terminated dendrimers in the almost fully ionized state also found higher effective charge densities for a lower generation $G_2$ than for $G_5$.[115] However, $\sigma_{eff}$ saturates to a fixed value for higher generations $G_5$ and $G_6$.[68]

In their CG simulations at 10 mM salt concentration, Xu et al.[68] found that $r_{eff}$ and $\sigma_{eff}$ depend weakly on salt concentration, which allows us to compare them with our simulations at $\rho_0$ = 25 mM salt concentration. Within a reasonable error, both $r_{eff}$ and $\sigma_{eff}$ evaluated from previous CG simulations by Xu et al.[68] are in good agreement with the other approaches used in our work. An exception is seen for the case of $G_0$ where charge renormalization has not been observed in the CG simulations. We also notice that the trend in $\sigma_{eff}$ with respect to the generation found in the CG simulations from Xu et al. is in agreement with the results of this study. Another consequence of the charge renormalization is the weak dependence of $\phi(r_{eff})$ with generation, which is observed in all approaches.

**4.2.5 Comparison of implicit- and explicit-water routes to the potential.** Fig. 5 shows the comparison of the electrostatic potentials from implicit- and explicit-water integration approaches. The long-range electrostatic potential obtained from the explicit-water approach also exhibits the long-range DH behaviour but with more statistical fluctuations and visible deviations from strict linearity. The slope, i.e., the inverse Debye length, is in most cases consistently close to the expected $\kappa$ = 0.52 nm$^{-1}$, corresponding to $\rho_0$ = 25 mM. We find that the occurring wiggles and deviations are caused by the large water fluctuations in the far field regime, rendering the integration prone to significant errors (see also the electrostatic fields shown in the ESI†). The curves from the explicit-route seem consistently shifted to larger distances by about the size of one water molecule, i.e., by about 0.3 nm.

While the values of $r_{eff}$ from the implicit approaches are not contradicting the explicit-route curves, no meaningful comparison can be made on a quantitative level. We conclude that, after all, all approaches give consistent values for the effective charge and size but within an uncertainty window of the size of one water molecule. More accurate quantifications are probably not meaningful to attempt, as they are obviously hampered by systematic uncertainties induced by continuum assumptions and micro-scale fluctuations.

### 4.3 Thermodynamic signature of dPGS–water interaction

Finally, we calculate the thermodynamic signature of dPGS–water association by calculating the temperature dependence of the water PMF.[69,70] As an illustrative example, a system of $G_2$-dPGS in 25 mM NaCl aqueous solution was simulated at two different temperatures, 283 K and 310 K, while keeping all other parameters constant and the simulation protocol the same (see Section 2). The PMFs of the dPGS–water interaction $V_w(r)$ are evaluated using eqn (1) for the two temperatures. Using the finite differences in eqn (11), the entropy profile $S_w(r)$ is evaluated, whereas the enthalpy is determined using eqn (12) at the mean temperature 296.5 K. Fig. 6 shows the distance-resolved profiles of the free energy in the form of the water PMF $V_w(r)$. Monotonically increasing PMFs with decreasing distance to the dPGS-COM indicate a net repulsion between dPGS and water. The corresponding changes in the system enthalpy $H_w$ and the entropic term $-TS_w$ of the dPGS hydration are also displayed. They show at first an increase of $H_w$ (and decrease of $-TS_w$) until ∼1 nm when approaching from larger distances. Since this location corresponds to the high density of terminal sulphate groups (see Fig. 2), this observation could be attributed to their ion-specific (Hofmeister) effects due to their chaotropic nature.[116–120] Unlike the kosmotropic divalent sulphates, the monovalent sulphates exhibit weaker interactions with water than water with itself and thus disturb the hydrogen-bond network of the surrounding water. Consequently, this leads to water with higher configurational freedom and thus higher entropy, simultaneously resulting in an enthalpic penalty for breaking water–water hydrogen bonds. The increase of $H_w$ until ∼1 nm is then followed by a rapid exchange of favourable/unfavourable compensating components at ∼0.6 nm. The unfavourable $V_w$ is dominated by the entropic term with a counterbalancing enthalpy $H_w$. The molecular origin of this effect could be credited to the dPGS interior environment rich with dPGS–oxygens, which is favourable for additional hydrogen bonds. However, simultaneously a steric hindrance of the dPGS core towards water and possibly localised dPGS–water hydrogen bond formation results in an entropic penalty. Ultimately, the entropic contribution dominates the dPGS–water interaction in the interior, consequently resulting in an unfavourable free energy. Surprisingly, despite a significant chemical difference of the dendrimers, very similar signatures were observed in explicit-water simulations of the PAMAM dendrimer.[21]

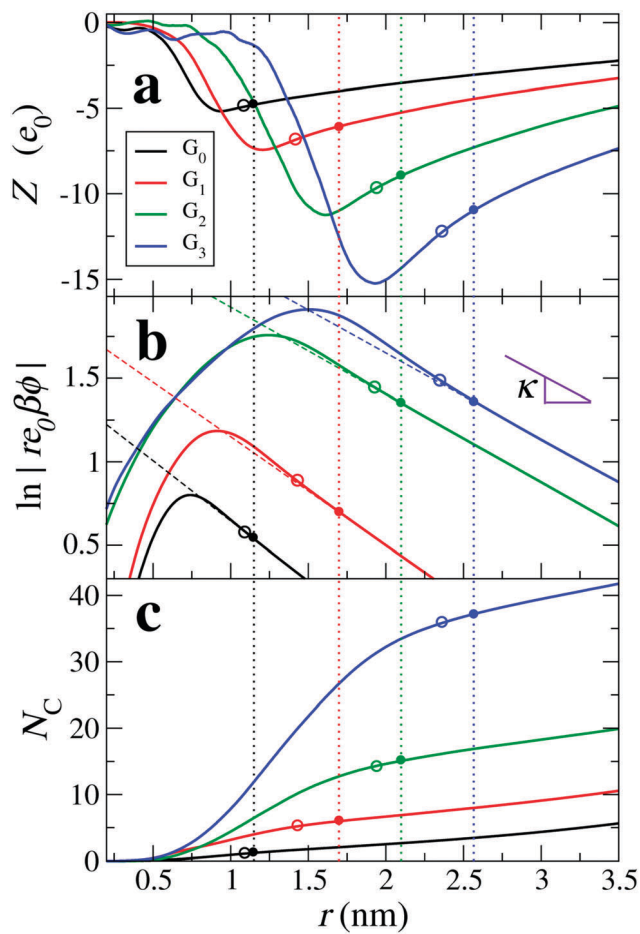

**Fig. 4** Results of the "implicit water" approach (including charge density distributions of all species except water and assuming $\varepsilon_r = 72$), (a) the net cumulative charge $Z(r)$ as a function of distance $r$ from the dPGS-COM, (b) the logarithm of the rescaled electrostatic potential, and (c) the cumulative counterion coordination $N_C(r)$. The dashed lines are fits of eqn (4) to the MD results in the far-field for the DH mapping (see Section 4.2.2). The vertical dotted lines denote the effective radii of dPGS, $r_{eff}$, intersecting the curves in filled circles. Empty circles denote radii $r_{inf}$ according to the inflection point criterion. The two circle types (corresponding to the DH mapping and the inflection point criterion) thus indicate the effective charge $Z_{eff}$, the effective potential $\phi(r_{eff})$ and the number of condensed counterions $N_C(r_{eff})$ on the respective vertical axes in (a), (b), and (c), respectively.

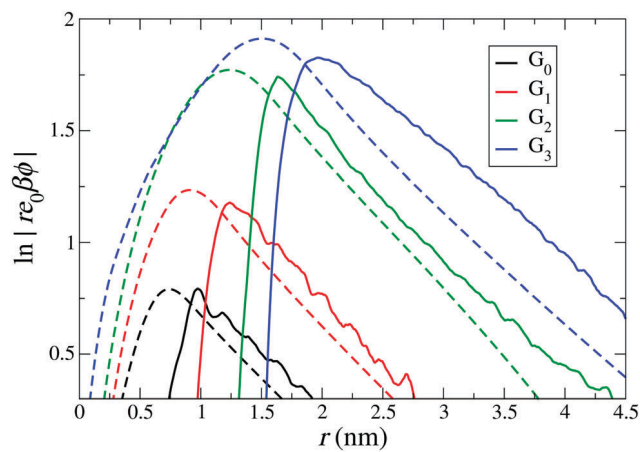

**Fig. 5** Comparison of electrostatic potential profiles with respect to distance from the dPGS-COM from the implicit and explicit water approaches, denoted by dashed and solid lines, respectively. The implicit water [the same as Fig. 4(b)] only takes dPGS and ion charges in a uniform dielectric medium ($\varepsilon_r = 72$) into account, while the explicit water approach additionally includes the partial charges of water in the vacuum permittivity.

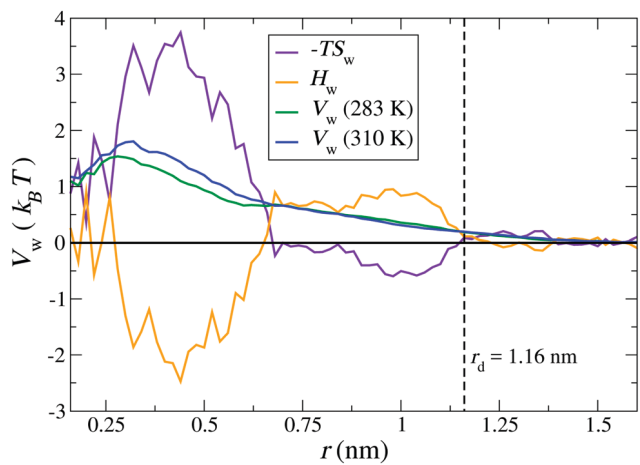

**Fig. 6** Distance-resolved thermodynamic profiles for water binding to the $G_2$-dPGS. The potential of mean force $V_w(r)$ is shown at 283 K and 310 K. The entropic profile $-TS_w$ and the enthalpy profile $H_w$ were calculated from the two PMFs by a finite difference derivative [eqn (11)] and the standard thermodynamic relation, eqn (12), at the mean temperature 296.5 K. The dashed vertical line indicates the bare radius $r_d$ of $G_2$-dPGS.

## 5 Conclusion

In summary, we have conducted explicit-water molecular dynamics simulations of dendritic polyglycerol sulphate, which is a biomedically important dendrimer and can be viewed as a representative of a class of highly charged dendritic polyelectrolytes. Beyond some general characterization of the ionic and hydration structure, in particular an electrostatic (surface) characterization of the dPGS was conducted in terms of the determination of the effective charge, the effective radius, and the surface potential by a direct mapping procedure of the calculated profiles onto the long-range Debye–Hückel electrostatic decays as well as using the inflection point criterion. By comparing these several routes to each other, but also implicit versus explicit-routes of integration towards the electrostatic potential, we found very consistent values for the effective charge size (with respect to far-field DH behaviour) within the uncertainty of the size of a water molecule.

Importantly, we can conclude that the coarse-grained models developed for the highly charged dPGS with explicit ions[68] are quite accurate from the electrostatic point of view. The latter can thus serve in future simulations and interpretations of the dPGS's and related dendritic polyelectrolytes' action in a biological context (e.g., interacting with proteins[10] or membranes) to understand and optimize their proven selective binding properties and efficacy in the medical treatment of inflammatory diseases.

## Conflicts of interest

There are no conflicts to declare.

## Acknowledgements

The authors thank Richard Chudoba for programming support and fruitful discussions. JD and MB acknowledge inspiring discussions with Qidi Ran and Rainer Haag. XX acknowledges funding from the Chinese Scholar Council (CSC). This project has received funding from the European Research Council (ERC) under the European Union's Horizon 2020 research and innovation programme (grant agreement No. 646659). The simulations were performed with resources provided by the North-German Supercomputing Alliance (HLRN).